\documentclass[prapplied,onecolumn,preprintnumbers,amsmath,amssymb,longbibliography]{revtex4-1}
\usepackage{graphicx}
\usepackage{color}      
\usepackage{xcolor} 
\usepackage{dcolumn}    
\usepackage{bm}         
\usepackage[caption=false]{subfig}
\usepackage{float}
\usepackage{relsize}
\usepackage{upgreek}
\usepackage{amsmath}
\usepackage{amssymb}
\usepackage{amsthm}

\begin{document}
\title{Experimental realization of the active convolved illumination imaging technique for enhanced signal-to-noise ratio}
\author{Wyatt Adams$^1$}
\author{Anindya Ghoshroy$^2$}
\author{Durdu \"O. G\"uney$^3$}
\email{Corresponding author: dguney@mtu.edu}
\affiliation{$^1$Ansys, Inc., 2600 Ansys Dr, Canonsburg, PA 15317, USA}
\affiliation{$^2$Caltech Optical Imaging Laboratory, Andrew and Peggy Cherng Department of Medical Engineering and Department of Electrical Engineering, California Institute of Technology, 1200 E. California Blvd., Pasadena, CA 91125, USA}
\affiliation{$^3$Department of Electrical and Computer Engineering , Michigan Technological University, 1400 Townsend Dr, Houghton, MI 49931-1295, USA}
\date{\today}

\begin{abstract}
Imaging is indispensable for nearly every field of science, engineering, technology, and medicine. However, measurement noise and stochastic distortions pose fundamental limits to accessible spatiotemporal information despite impressive tools such as SIM, PALM/STORM, and STED microscopy. How to combat this challenge ideally has been an open question for decades. Inspired by a `\textit{virtual gain}' technique to compensate losses in metamaterials, `\textit{active convolved illumination}' has been recently proposed to significantly improve the signal-to-noise ratio, hence data acquisition. In this technique, the light pattern of the object is superimposed with a correlated auxiliary pattern, the function of which is to reverse the adverse effect of noise and random distortion based on their spectral characteristics. Despite enormous implications in statistics, an experimental realization of this novel technique has been lacking to date. Here, we present the first experimental demonstration. We find that the active convolved illumination does not only boost the resolution limit and image contrast, but also the resistance to pixel saturation. The results confirm the previous theories and opens up new horizons in a wide range of disciplines from atmospheric sciences, seismology, biology, statistical learning, and information processing to quantum noise beyond the fundamental boundaries.
\end{abstract}
\maketitle

\section{Introduction}
Imaging is an indispensable tool in the toolbox of nearly every field of science, engineering, technology, and medicine. Unfortunately, encoding the desired information into electromagnetic waves imposes a limit to the performance of imaging systems at the outset -- the detection of the fields by the interaction of photons (the light signal) and matter (the light detector) means that the signal-to-noise ratio (SNR) for long exposures will always be limited physically by shot noise. A na\"ive analysis would reveal that adding up more photons in the detector would lead to higher SNR. This is true, however typically (e.g., for incoherent light), the magnitude of the transfer function for an imaging system with an unobstructed pupil decreases with increasing spatial frequency \cite{goodman_introduction_2005}. It follows that the spectral SNR then also decreases with increasing spatial frequency, since the shot noise variance is constant in the spatial frequency domain \cite{ghoshroy2020theory,ingerman_signal_2019,becker_better_2018}. Consequently, adding up more photons blindly does not lead to much increase for the SNR of high spatial frequencies. One also does not have the freedom to arbitrarily increase the number of photons collected, since at some point the detector will become saturated. Each specific imaging modality will have its specific limitations. For example, in fluorescence imaging only a certain exposure can be obtained before photobleaching occurs and high intensity becomes detrimental for live specimen.

Subwavelength optical engineering through metamaterials and metasurfaces offer unprecedented opportunities in a wide range of applications such as superresolution imaging \cite{pendry2000negative,adams2016review,ponsetto2017experimental,ma2017super,bezryadina2018high,ma2018experimental,lee2021metamaterial}, photolithography \cite{gao2015enhancing,liang2018achieving}, wireless and optical communications \cite{si2019broadband,yang2020tunable}, multifunctional and flat optics and photonics \cite{yu2014flat,zhou2019multifunctional,zhou2020flat,wan2019visible,rho2020metasurfaces}, metalenses \cite{banerji2019imaging}, intelligent metaphotonics \cite{krasikov2021intelligent}, light detection and ranging \cite{lesina2020tunable}, autonomous vehicles \cite{zecca2019symphotic}, and quantum information \cite{asano2015distillation,uriri2018active,mirhosseini2018superconducting}. However, the photon losses hinder their further viability \cite{khurgin2010search,krasnok2020active,ghoshroy2020loss}. Inspiration from research in loss compensation for metamaterials and plasmonics employing `\textit{virtual gain}' \cite{potma2003picosecond,sadatgol2015plasmon,li2020virtual,ghoshroy2020loss} led us to propose a unique perspective on the noisy imaging problem \cite{sadatgol2015plasmon,adams_bringing_2016,zhang_enhancing_2016,zhang_analytical_2017,ghoshroy_active_2017,adams_plasmonic_2017,adams_plasmonic_2018,ghoshroy_hyperbolic_2018,ghoshroy_enhanced_2018,ghoshroy2020theory}. The fundamental resolution limit to superresolving lenses is not determined by the diffraction limit, but rather by a shot noise limit, i.e. where the shot noise overcomes the transfer function in the spatial frequency domain \cite{wang2019deep,cox1986information,katznelson2004introduction,becker_better_2018}. How to tackle this problem ideally has been an open question for decades \cite{cox1986information,katznelson2004introduction,roggemann1992linear,biemond1990iterative,richardson1972bayesian,lucy1974iterative}. The \textit{active convolved illumination} (ACI) technique has recently been proposed theoretically as a ubiquitous noise and distortion mitigation scheme to improve the image signal-to-noise (SNR) ratio by systematically manipulating the image spectrum depending on the underlying stochastic behavior \cite{sadatgol2015plasmon,ghoshroy_active_2017,adams_plasmonic_2018,ghoshroy_hyperbolic_2018,ghoshroy_enhanced_2018,ghoshroy2020theory}. Popular techniques to improve image resolution or SNR are structured illumination microscopy (SIM) \cite{schermelleh2019super,kner2009super,schermelleh2008subdiffraction,york2012resolution,york2013instant,ingerman_signal_2019}, stimulated emission depletion (STED) microscopy \cite{schermelleh2019super,wegel2016imaging,bottanelli2016two}, stochastic optical reconstruction microscopy/photoactivated localization microscopy (STORM/PALM) \cite{schermelleh2019super,jones2011fast,takakura2017long}, and computational methods \cite{cox1986information,katznelson2004introduction,roggemann1992linear,biemond1990iterative,richardson1972bayesian,lucy1974iterative,bertero2003super,zeng2017computational,xing2020computational} including machine learning \cite{friedman2001elements,zaknich2005principles,wang2019deep}. An interesting method for far-field imaging with a constant ``photon budget'' (i.e., a constant number of photons in the object plane), based on a split-pupil-optimization, has recently been put forward to break the SNR limit imposed by Fermat's principle \cite{becker_better_2018}. However, ACI operates down at the physical layer and enhances data acquisition, thus expected to benefit both conventional and novel approaches.

\begin{figure}
    \centering
    \includegraphics[width=\textwidth]{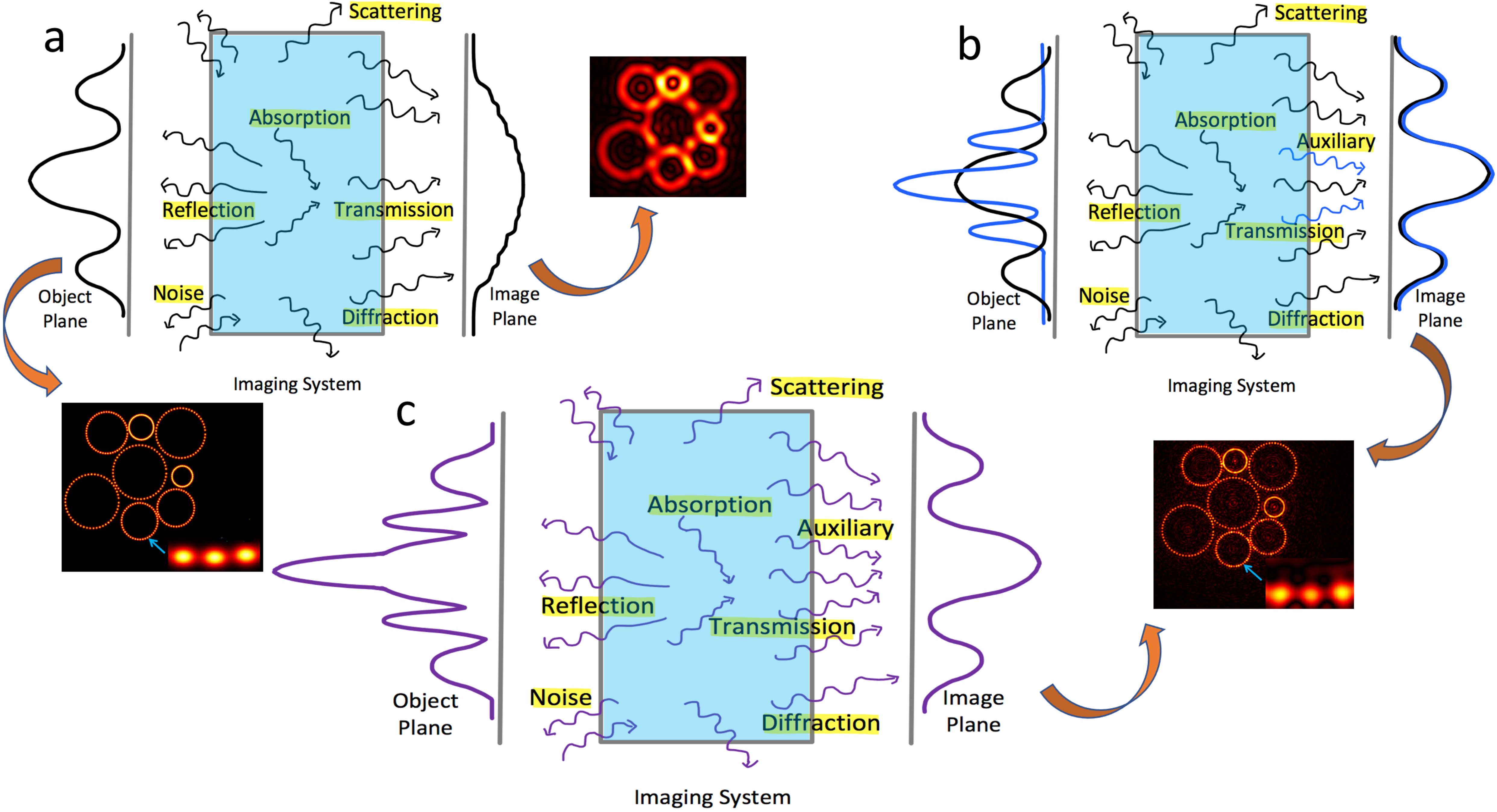}
    \caption{Working principle of the active convolved illumination technique. (a) Most imaging systems suffer from loss of information due to physical processes such as reflection, absorption, scattering, diffraction, and noise. This results in an image with a poor resolution and contrast. (b) In the ACI technique, the light pattern forming the object (black line) is superimposed with an auxiliary pattern (blue line), the purpose of which is to compensate the adversary photons that cause the information leakage from the system. The auxiliary is correlated with the object and based on the underlying spectral noise distribution. The result is a remarkable acquisition of data, hence high-fidelity image. (c) Equivalent to (b) except that the auxiliary and object patterns are implemented as a single entity as in this work.}
    \label{fig:ACI}
\end{figure}

Fig. \ref{fig:ACI} illustrates the working principle of the ACI technique. As shown in Fig. \ref{fig:ACI}(a), most imaging systems suffer from various mechanisms of signal photon loss such as impedance mismatch, absorption, scattering, diffraction, and noise. This, in general, results in a low-fidelity information transfer from the object plane to the receiver plane. It was hypothesized that the ACI could overcome the loss of information as depicted in Fig. \ref{fig:ACI}(b) \cite{sadatgol2015plasmon,ghoshroy_active_2017,adams_plasmonic_2018,ghoshroy2020theory}. The light pattern which forms the object (black line) is superimposed with an auxiliary pattern (blue line) correlated with the object. Therefore, the light pattern on the object plane differs from the actual object. The purpose of the auxiliary light is to compensate the adversary photons, so that the object is transferred through the system unscathed. The auxiliary pattern is typically found by characterizing the spectral distribution of noise obtained from the reference imaging system as the noise poses the fundamental limit to accessible spatial information \cite{ghoshroy2020theory}. The auxiliary and object patterns need not to be two separate entities but can also be implemented as a single entity as shown in Fig. \ref{fig:ACI}(c). Also, the negative values for the auxiliary is not precluded. For incoherent imaging this physically means the energy is reduced at those locations, which is indeed useful to prevent excessive noise, long average exposure, and pixel saturation.

The previous works on the foundations of ACI \cite{sadatgol2015plasmon,ghoshroy_active_2017,adams_plasmonic_2018,ghoshroy_hyperbolic_2018,ghoshroy_enhanced_2018,ghoshroy2020theory}, along with further inspiration from other research in far-field imaging \cite{becker_better_2018,ingerman_signal_2019}, led us to construct the current work, which presents the first experimental realization of ACI. Following the implementation sketched in Fig. \ref{fig:ACI} (c), we find for incoherent light that by suppressing the detection of the high-SNR spatial frequency harmonics in an object, while amplifying the magnitude of those with low SNR, the resulting image can have large SNR for previously low-SNR spatial frequencies, due to a prudent control of the flat overall noise level dictated by the Poisson distribution. This does not only manage the exposure time  efficiently, but also leads to enhanced image resolution and contrast beyond what is possible with common post-processing.

The previous body of literature \cite{sadatgol2015plasmon,ghoshroy_active_2017,adams_plasmonic_2018,ghoshroy_hyperbolic_2018,ghoshroy_enhanced_2018,ghoshroy2020theory} on ACI have been only theoretical and typically utilized a high spatial frequency passband function, in conjunction with an increased exposure, to enhance the resolution performance of thin metal films acting as near-field plasmonic ``superlenses" \cite{ghoshroy_active_2017,adams_plasmonic_2018,ghoshroy_hyperbolic_2018,ghoshroy_enhanced_2018,ghoshroy2020theory}. This passband function was used to not only compensate the losses inherent in the metal film (i.e., virtual gain), but also to improve the image spectrum SNR similar to what we have shown in this work at far-field. This imaging method has been called \cite{adams_plasmonic_2018} `\textit{active convolved illumination}' for a couple reasons. First, the term `active' was chosen since an added energy \cite{ghoshroy_hyperbolic_2018} was used to obtain enhancement over the control (a bare superlens). Secondly, the term `convolved' was chosen since the applied passband function is physically convolved with the fields emanating from the object. For coherent illumination, this passband function can be realized by a hyperbolic metamaterial (HMM) \cite{ghoshroy_hyperbolic_2018,ghoshroy_enhanced_2018}. While there is no Fourier plane in the near-field configuration, the HMM can modify the Fourier components of the incident evanescent waves by its dispersion. In this case, the pupil function can then simply be thought of as the coherent transfer function of the cascaded HMM-superlens system and the SNR for the large spatial frequencies is enhanced with respect to the control.

Below, first a theory of ACI for incoherent light is presented that shows how to improve the SNR for the high spatial frequencies (i.e., low-SNR components) of an image obtained from an object illuminated with incoherent light. The theory is implemented in numerical simulations to predict the resolution enhancement, and experimental images are collected using a low numerical aperture imaging system to confirm the predictions. 
The end result is an image with higher resolution and improved contrast as compared to the control image. It is also shown that ACI can help prevent pixel saturation for longer exposures. The experimental work here confirms that the theories of ACI are, in fact, both consistent with real noisy optical signals, and can be straightforwardly extended to conventional far-field imaging systems and potentially to complex systems with different noise and distortion characteristics. A detailed understanding of random processes in the Fourier domain, crucial for the implementation of the ACI, and the subsequent development of the auxiliary pattern (see Fig. \ref{fig:ACI}(b)) or the equivalent combined light pattern for the object (see Fig. \ref{fig:ACI}(c)), immune to such distorting effects, may in general open multiple avenues of research. It may lead to a generalized theory of randomly distorted systems pervading a wide range of disciplines from atmospheric sciences, seismology, and biology to the mesoscopic physics of noisy quantum systems.

\section{Theoretical Model}\label{section:Chapter5-Theory}
\subsection{Incoherent Imaging}
Consider a uniform beam of spatially incoherent, narrowband light with photon flux density $\Phi_0$ [$\mathrm{photons}/m^2 \cdot s$] striking a planar, transmissive object. After passing through the object, the light has spatial variations and the resulting transmitted photon flux density is given by $O(\mathbf{r})$, with $\mathbf{r}\in\mathbb{R}^2$ denoting the position coordinate on the plane. The process of mapping this light distribution with an imaging system can be represented by a convolution
\begin{equation}\label{eq:imaging}
    I(\mathbf{r})=H(\mathbf{r})*O(\mathbf{r}),
\end{equation}
where $I(\mathbf{r})$ is the photon flux density on the image plane, $H(\mathbf{r})$ is the point spread function of the imaging system, and $*$ denotes the convolution. Here it should be noted that $I(\mathbf{r})$, $O(\mathbf{r})$ $\in\mathbb{R}_{\geq0}$ since they represent flux densities and not complex fields. Applying the convolution theorem and using the normalized image and object spectra, $\tilde{I}(\mathbf{k})$ and $\tilde{O}(\mathbf{k})$, respectively, Eq. \ref{eq:imaging} gives
\begin{equation}\label{eq:imaging-ft}
    \tilde{I}(\mathbf{k})=\tilde{H}(\mathbf{k})\tilde{O}(\mathbf{k}),
\end{equation}
where $\tilde{H}(\mathbf{k})$ is the optical transfer function (OTF) of the imaging system \cite{goodman_introduction_2005}. In general $\tilde{I}(\mathbf{k})$, $\tilde{H}(\mathbf{k})$, $\tilde{O}(\mathbf{k})\in\mathbb{C}$. For later use, we define $|\tilde{H}(\mathbf{k})|$ as the modulation transfer function (MTF).

\subsection{Detection and Noise}
Eqs. \ref{eq:imaging} and \ref{eq:imaging-ft} assume continuous signals in position and reciprocal space. To incorporate practical detection of the deterministic signal $I(\mathbf{r})$, let us consider the case where we collect an image on the image plane using a photoelectric detector with $n_p$ pixels. The center of the $p^\mathrm{th}$ pixel ($p \in \mathbb{Z}_+$) is at $\mathbf{r}_p=(x_i,y_j)$, and the pixels are rectangular with side lengths $\Delta x$ and $\Delta y$ along the $x$ and $y$ dimensions, respectively. Then, at the $p^\mathrm{th}$ pixel, the expected number of detected photons is given by
\begin{equation}\label{eq:photons-at-pth-pixel}
    \begin{split}
        \bar{I}_p & = \eta T \int_{A_p} I(\mathbf{r}) d^2\mathbf{r} \\
        & = \eta T \int_{y_j - \Delta y/2}^{y_j + \Delta y/2} \int_{x_i - \Delta x/2}^{x_i + \Delta x/2} I(x,y) dx dy \\
        & \approx \eta T \Delta x \Delta y I(x_i,y_j),
    \end{split}
\end{equation}
where $\eta$ is the quantum efficiency of the pixel and $T$ is the exposure time or integration time. The approximation in the third line of Eq. \ref{eq:photons-at-pth-pixel} assumes that the signal $I(\mathbf{r})$ is slowly varying across the area of the pixel, i.e. the signal is well sampled.
Since Eq. \ref{eq:photons-at-pth-pixel} describes a sampling of a spatial distribution of discrete particles (photons), there will be an intrinsic randomness due to shot noise in the photon signal $I_{p,\gamma}$. In this case, the probability mass function (PMF) is
\begin{equation}\label{eq:Poisson}
    \mathcal{P} \left(I_{p,\gamma}|\bar{I}_p \right) = \frac{\left( \bar{I}_p \right)^{I_{p,\gamma}}}{I_{p,\gamma}!}e^{-\bar{I}_p}.
\end{equation}
This is of course coming from the fact that the counting of discrete particles at a constant rate follows a Poisson distribution, for which
\begin{equation}\label{eq:shot-noise-variance}
    \mathrm{Var}(I_{p,\gamma})=\bar{I}_p.
\end{equation}

In most photoelectronic imaging detectors, such as complementary metal-oxide-semiconductor (CMOS) or charge-coupled device (CCD) cameras, there are primarily two sources of noise. The first is due to the statistics of Eq. \ref{eq:Poisson}, the shot noise which is dependent on the photon signal. The second is noise from the readout electronics, which is independent of the photon signal. We can then write the detected signal as
\begin{equation}\label{eq:noise-addition}
    I_p=\bar{I}_p+N_{p,\gamma}+N_{p,e},
\end{equation}
where $N_{p,\gamma}$ is a discrete random variable representing the shot noise with PMF described by Eq. \ref{eq:Poisson}, and $N_{p,e}$ is a discrete random variable representing the electronic readout noise.

In order to show how the noise addition in Eq. \ref{eq:noise-addition} affects the image spectrum, we compute the discrete Fourier transform
\begin{equation}\label{eq:pixel-value-dft}
    \begin{split}
        \tilde{I_q}&=  \sum_p I_p e^{-i2\pi \mathbf{k}_q \cdot \mathbf{r}_p} \\
        &=  \sum_p (\bar{I}_p+N_{p,\gamma}+N_{p,e}) e^{-i2\pi \mathbf{k}_q \cdot \mathbf{r}_p} \\
        &=   \sum_p \bar{I}_p e^{-i2\pi \mathbf{k}_q \cdot \mathbf{r}_p} + \sum_p N_{p,\gamma} e^{-i2\pi \mathbf{k}_q \cdot \mathbf{r}_p} + \sum_p N_{p,e} e^{-i2\pi \mathbf{k}_q \cdot \mathbf{r}_p}  \\
        &=   \tilde{\bar{I}}_q + \tilde{N}_{q,\gamma} + \tilde{N}_{q,e},
    \end{split}
\end{equation}
where $q \in \mathbb{Z}_+$ and $\{\mathbf{k}_q=(k_{x,l},k_{y,m}) \mid 1 \leq q \leq n_p\}\subset\{\mathbf{k}\in\mathbb{R}^2\}$ is the Fourier space corresponding to the pixelated position space \{$\mathbf{r}_p \mid 1 \leq p \leq n_p$\}. We then consider the statistical properties of $\tilde{N}_{q,\gamma}$ and $ \tilde{N}_{q,e}$. Since the shot noise variance is known from Eq. \ref{eq:shot-noise-variance} (replacing $I_{p,\gamma}$ with $N_{p,\gamma}$) and we can assume the pixels are statistically independent, we can write \cite{ghoshroy2020theory,ingerman_signal_2019,becker_better_2018}
\begin{align}\label{eq:Fourier-shot-noise-variance}
    \mathrm{Var}(\tilde{N}_{q,\gamma}) &=\mathrm{Var} \left (\sum_p N_{p,\gamma}e^{-i2\pi\mathbf{k}_q\cdot\mathbf{r}_p} \right )=\sum_p \mathrm{Var}(N_{p,\gamma})|e^{-i2\pi\mathbf{k}_q\cdot\mathbf{r}_p}|^2 
    =\sum_p \bar{I_p}=n_\gamma,
\end{align}
where $n_\gamma$ is the total expected number of photons in the entire image. In words, Eq. \ref{eq:Fourier-shot-noise-variance} states that the variance in Fourier space is constant, and is controlled by the total expected number of photons collected on the detector. We can write a similar equation for the readout noise \cite{ingerman_signal_2019},
\begin{equation}\label{eq:Fourier-read-noise-variance}
    \mathrm{Var}(\tilde{N}_{q,e})=\mathrm{Var} \left (\sum_p N_{p,e}e^{-i2\pi\mathbf{k}_q\cdot\mathbf{r}_p} \right )=\sum_p \mathrm{Var}(N_{p,e})|e^{-i2\pi\mathbf{k}_q\cdot\mathbf{r}_p}|^2=\sum_p \sigma_{p,e}^2,
\end{equation}
where $\sigma_{p,e}^2$ is the readout noise variance at pixel $p$ and again the assumption is made that the pixels are statistically independent. Let us also assume that $\sigma_{p,e}^2 = \sigma_e^2$, meaning every pixel has similar electrical performance in terms of noise. Then Eq. \ref{eq:Fourier-read-noise-variance} becomes
\begin{equation}
    \mathrm{Var}(\tilde{N}_{q,e}) = n_p \sigma_e^2.
\end{equation}
Therefore, the spectral readout noise variance is also a constant, and scales linearly with the number of pixels. For modern cameras, the readout noise is usually minimal such that it is neglected, though we keep it here for completeness.

\subsection{Tailoring the Spectral SNR}
From imaging theory, we know that the optical transfer function of an incoherent imaging system is given by the normalized autocorrelation \cite{goodman_introduction_2005} of the system's pupil function $\tilde{P}(\mathbf{k})$, or
\begin{equation}\label{eq:incoherent-OTF-from-pupil}
    \tilde{H}(\mathbf{k})= \frac{\int \tilde{P}(\bm{\kappa})\tilde{P}^*(\bm{\kappa}-\mathbf{k})d^2\bm{\kappa}}{\int |\tilde{P}(\bm{\kappa})|^2 d^2\bm{\kappa}}.
\end{equation}
In the discrete notation described in the previous section, we can write
\begin{equation}\label{eq:discrete-autocorrelation}
    \tilde{H}_q= \frac{ \sum_{\bm{\kappa}} \tilde{P}_{\bm{\kappa}}\tilde{P}^*_{\bm{\kappa}-\mathbf{k}_q} }{ \sum_{\bm{\kappa}} |\tilde{P}_{\bm{\kappa},o}|^2}=\alpha\tilde{P}_q\star \tilde{P}_q,
\end{equation}
where $\tilde{P}_{\bm{\kappa},o}$ is a reference pupil, $\alpha$ is the normalization constant, and $\tilde{P}_q$ is the pupil function in the discrete notation. To normalize \ref{eq:discrete-autocorrelation}, the reference pupil is conventionally chosen as the pupil itself, making the DC pixel of $\tilde{H}_0=1$, similar to Eq. \ref{eq:incoherent-OTF-from-pupil}. However we define a reference pupil in Eq. \ref{eq:discrete-autocorrelation} in order to later directly compare two different pupil functions. Consider an incoherent imaging system in air with maximum resolvable spatial frequency $k_{max}=2\text{NA}k_0$, where NA is the numerical aperture, $k_0=1/\lambda_0$, and $\lambda_0$ is the center free space wavelength of the illumination source. The pupil function is assumed to have circular symmetry about the optical axis (which is along $z$-direction) and we define it as
\begin{equation}\label{eq:pupil-function}
    \tilde{P}_q=
    \begin{cases}
        1 &\text{if } k_- \leq |\mathbf{k}_q| \leq k_+\\
        0 &\text{otherwise},
    \end{cases}
\end{equation}
where  $k_- \geq 0$ and $k_- < k_+ \leq k_{max}/2$. From Eq. \ref{eq:pupil-function} we can see that setting $k_-=0$ and $k_+=k_{max}/2$ gives a typical diffraction-limited imaging system with open pupil. We choose this case as our reference pupil $\tilde{P}_{q,o}$. However, if we make $k_-$ nonzero, we introduce an obstruction in the central portion of the pupil, which has the effect of lowering the overall transmission with respect to the reference pupil, and also preferentially reinforcing larger spatial frequencies with respect to the smaller ones in comparison to the reference pupil.

An important metric for an imaging system is its ability to discern image spectrum content from noise, or its spectral SNR. To relate the pupil function to the spectral SNR, we first use Eqs. \ref{eq:imaging-ft} and \ref{eq:discrete-autocorrelation}, without normalization, to rewrite the expected image spectrum as
{\begin{equation}
    \tilde{\bar{I}}_q=\left[\beta \tilde{P}_q\star \tilde{P}_q\right]\tilde{O}_q=\tilde{\mathcal{H}}_q\tilde{O}_q,
\end{equation}}
where $\beta \tilde{P}_q\star \tilde{P}_q=\tilde{\mathcal{H}}_q$ is the unnormalized OTF and the constant $\beta$ has replaced $\alpha$ due to the fact that the OTF gives the system response with respect to $\mathbf{k}_q=0$ \cite{tyo2010linear}.
Using a standard definition of SNR, we then can write
\begin{equation}\label{eq:spectral-SNR}
    \mathrm{SNR}_q = \frac{|\tilde{\bar{I}}_q|}{\sqrt{n_\gamma+n_p\sigma_e^2}}=\frac{\left|\left[\beta \tilde{P}_q\star \tilde{P}_q\right]\tilde{O}_q\right|}{\sqrt{n_\gamma+n_p\sigma_e^2}}=\frac{\left|\tilde{\mathcal{H}}_q\tilde{O}_q\right|}{\sqrt{n_\gamma+n_p\sigma_e^2}}
\end{equation}
as the image spectrum SNR. In Eq. \ref{eq:spectral-SNR}, the numerator is the expected image spectrum, which can be engineered by manipulation of the pupil function, and the denominator is the total standard deviation of the signal from the photonic and electronic noise terms in Eqs. \ref{eq:Fourier-shot-noise-variance} and \ref{eq:Fourier-read-noise-variance}. An obvious consequence of Eq. \ref{eq:spectral-SNR} is that reducing $n_\gamma$ will decrease the constant noise floor in the image spectrum. The signal in the numerator will also decrease similarly, but can be engineered through $\tilde{P}_q$ to enhance different portions of the spectrum. Conversely, increasing $n_\gamma$ will increase the constant noise floor, but this will also increase the signal. If the readout noise is neglected in Eq. \ref{eq:spectral-SNR}, $\mathrm{SNR}_q$ will then theoretically increase with $\sqrt{T}$ (see Eq. \ref{eq:photons-at-pth-pixel}). Interestingly, if one can additionally manipulate the pupil function to amplify certain spectral regions, further improvement in $\mathrm{SNR}_q$ becomes possible in those regions beyond original shot noise limited signal. Engineering the system to preferentially pass a certain band or bands of the image spectrum, while keeping the overall noise unchanged can also improve $\mathrm{SNR}_q$ for the selected regions. To put the latter scenario into analogy, we are given a ``photo-budget" (i.e., a constant number of photons in the image plane)  $n_\gamma$, and we can freely decide how to spend that budget in the image spectrum via $\tilde{P}_q$ so that we achieve an improved SNR for certain spatial frequencies. Boosting SNR with such spectrum manipulation scenarios above considering the underlying noise statistics forms the essence of ACI and is the central theme of this work. The relation of these processes with the auxiliary pattern can be explained as follows. Let  $\tilde{\mathcal{H}}_{q,o}$ and $\tilde{\mathcal{H}}_{q,A}$ be the unnormalized OTFs for the imaging systems without (i.e., reference) and with ACI, respectively. Then, the expected image spectrum obtained by the auxiliary component becomes
\begin{equation}
    \tilde{\bar{I}}_{q,aux}=\tilde{\mathcal{H}}_{q,A}\tilde{O}_q-\tilde{\mathcal{H}}_{q,o}\tilde{O}_q=\tilde{h}_q\tilde{\mathcal{H}}_{q,o}\tilde{O}_q-\tilde{\mathcal{H}}_{q,o}\tilde{O}_q=\tilde{\mathcal{H}}_{q,o}(\tilde{h}_q-1)\tilde{O}_q,
\end{equation}
where we define the virtual gain factor $\tilde{h}_q=\tilde{\mathcal{H}}_{q,A}/\tilde{\mathcal{H}}_{q,o}$. Therefore, the auxiliary pattern spectrum corresponding to Fig. \ref{fig:ACI} (b) and the combined object spectrum with embedded auxiliary corresponding to Fig. \ref{fig:ACI} (c) can be, respectively, expressed as \begin{align}
    \tilde{O}_{q,aux}&=(\tilde{h}_q-1)\tilde{O}_q,\\
    \label{eq:embedded_aux}
    \tilde{O}_{q,A}&=\tilde{h}_q\tilde{O}_q,
\end{align}
which are both correlated with the object. $\tilde{\mathcal{H}}_{q,A}$ in Eq. 16 is also known as the `\textit{active transfer function}' (ATF) \cite{ghoshroy2020theory} and proportional to the exposure time. If the ATF leads to amplification at some spatial frequencies with respect to the reference system, this implies virtual gain in that spectral region. The upcoming section describes a simple experimental construction of the embedded auxiliary used in this work.


\section{Experiment}

\begin{figure}
    \centering
    \includegraphics[width=\textwidth]{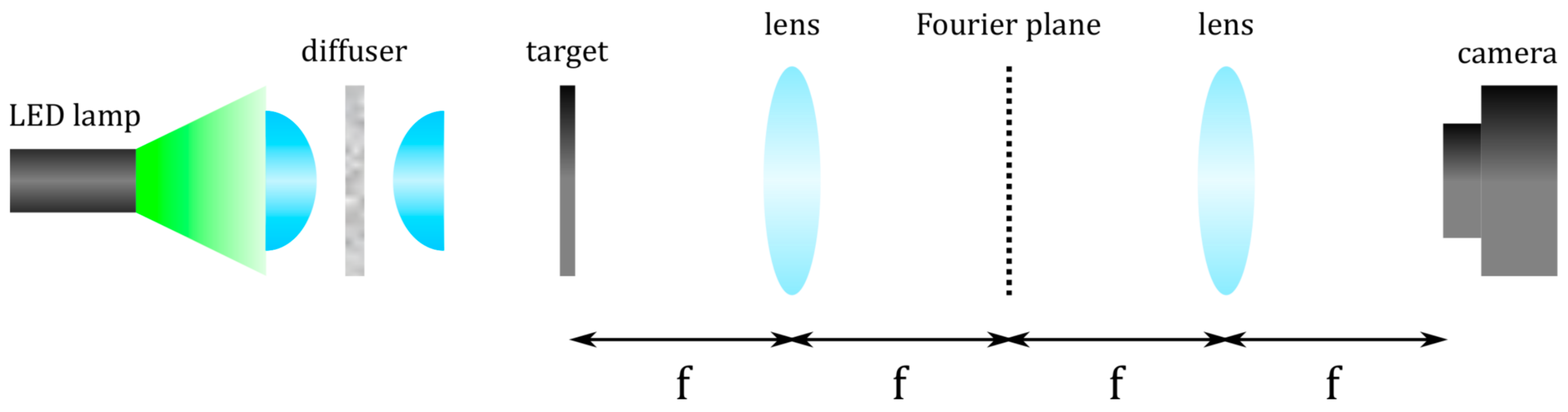}
    \caption{Imaging experiment configuration. A diffused LED source illuminates an imaging target, which is then processed by a 4$f$ system consisting of two achromatic doublet lenses and a transparency in the Fourier plane. The images are detected with a CMOS camera.}
    \label{fig:experiment-diagram}
\end{figure}

In order to implement the above spectral SNR manipulation into an experimental imaging system, we simply need access to the Fourier plane in order to manipulate the pupil function. Therefore, we chose to construct a typical 4$f$ system with no magnification as shown in Fig. \ref{fig:experiment-diagram} in order to simplify the analysis and experiment. The ACI here is then simply implemented with annular pupil and variable photon exposure, and the reference system with open pupil. However, it should be emphasized that the concepts presented should be generally applicable to any imaging system which is linear and shift-invariant, provided there is a mechanism with which to manipulate the Fourier content of the light.
\begin{figure}
    \centering
    \includegraphics[width=0.9\textwidth]{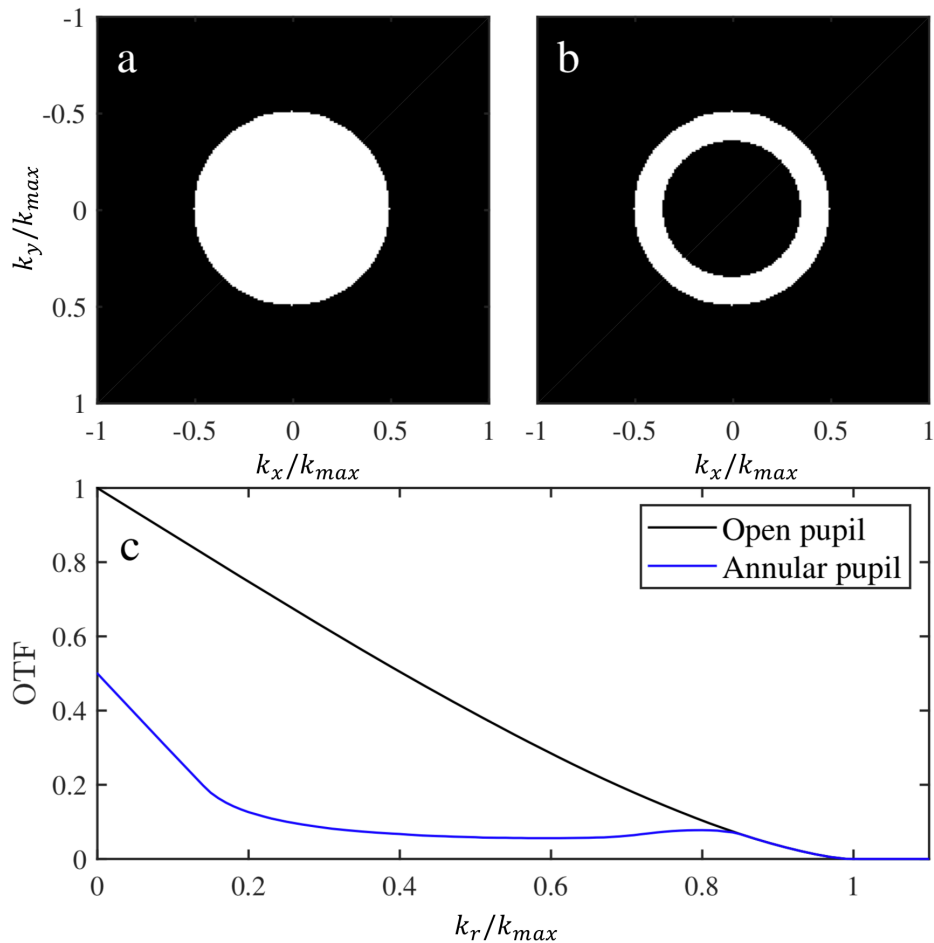}
    \caption{(a) Open (reference) and (b) annular pupils with (c) corresponding OTFs calculated from Eq. \ref{eq:discrete-autocorrelation} as a function of radial spatial frequency $k_r=\sqrt{k_x^2+k_y^2}$ used in the experiment in Fig. \ref{fig:experiment-diagram}. Both pupils have $k_+=k_{max}/2$. In (a), $k_-=0$ and in (b), $k_-=k_{max}/2\sqrt{2}$. In the actual experiment, we chose an outer pupil diameter of 5 mm for (a) and (b), making NA$=0.0066$. In (c) the OTF for the annular pupil was normalized with respect to open pupil to illustrate their relative transfer characteristics.}
    \label{fig:pupil-functions}
\end{figure}

The experiment in Fig. \ref{fig:experiment-diagram} images an object (target) illuminated by a narrowband incoherent light source (Thorlabs LIU525B). Before hitting the target, the light is focused onto a diffuser in order to decrease the spatial coherence and to avoid imaging the light source onto the Fourier plane. Then the light is roughly collimated by a second lens before hitting the target. The light distribution exiting the target is transferred through an achromatic-doublet lens (Space Optics Research Labs) with a focal length of $f=38.1$ cm. On the Fourier plane, a pupil transparency is placed that has either a circular or annular opening, as shown in Fig. \ref{fig:pupil-functions}. For all the images, we chose an outer pupil diameter of $d=5$ mm, making NA$=0.0066$ using the formula NA$=d/2f$. The transparencies were printed with a photoplotter onto transparent plastic sheets, then cut out and mounted in standard optical mounts. After passing through the pupil on the Fourier plane, the light is again transferred through a second identical achromatic-doublet which then focuses the resulting image onto a CMOS camera (Thorlabs DCC1645C).

The goal of this experiment was to show directly an enhancement in image resolution by modifying the pupil to improve $\mathrm{SNR}_q$ for the largest spatial frequencies in accordance with Eq. \ref{eq:spectral-SNR}. This led to the annular pupil configuration in Fig. \ref{fig:pupil-functions} (b). To show quantitatively the improvement in the resolution performance afforded by the annular pupil configuration over the open pupil, we replaced the resolution target with a 10 $\mu$m diameter pinhole. Since this diameter is below the diffraction limit for the chosen numerical aperture defined by $k_+$ and $f$, the resulting image of the pinhole gives the PSF of the imaging system. These PSF images were taken with both the open and annular pupil transparencies with varying exposure times. From these images, the transfer functions and corresponding $\mathrm{SNR}_q$ can be computed for each exposure. It should be noted that in the present implementation, the annular pupil that controls the image spectrum by varying the photon exposure, in effect, directly produces the required auxiliary in the Fourier domain as minted into the compound object-auxiliary pattern given by Eq. \ref{eq:embedded_aux}. The level of exposure and the annular pupil dimensions are determined carefully from the limiting spatial frequency of the reference system and the fact that \textit{shot noise variance is constant and equal to the total expected number of photons} (see Eq. \ref{eq:Fourier-shot-noise-variance}).


\subsection{Point Spread Functions and Transfer Functions}

\begin{figure}
    \centering
    \includegraphics[scale=0.33]{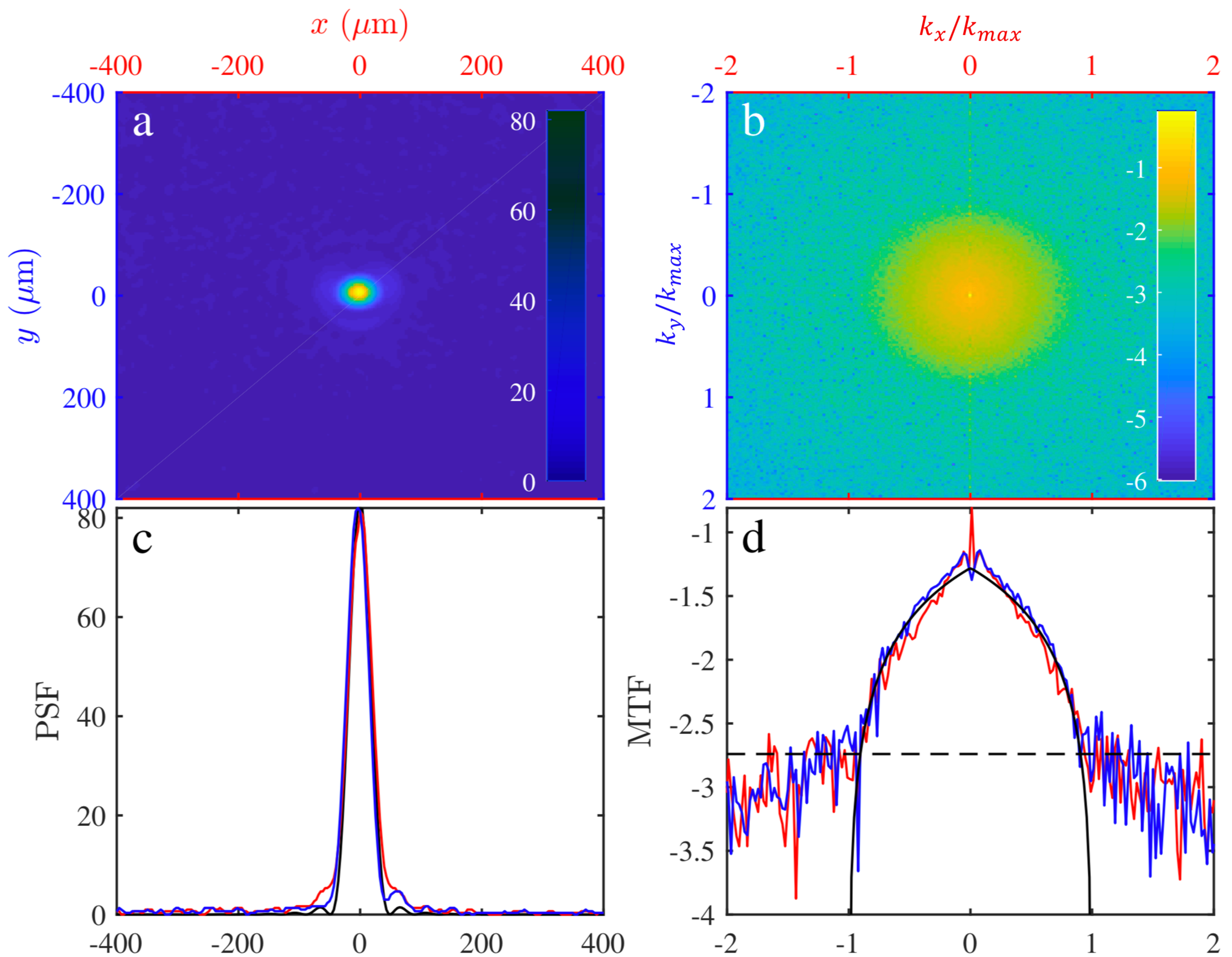}
    \caption{Measured PSF and MTF for full, unobstructed pupil with NA$=0.0066$ and exposure time of $T=2$ s. (a) The PSF collected from the setup in Fig. \ref{fig:experiment-diagram}. (b) The MTF calculated by fast Fourier transformation of (a). (c) Cross-sections of (a) through the origin along $x$ (red line) and $y$ (blue line). The theoretical prediction is given by the black line. (d) Cross-sections of (b) through the origin along $k_x$ (red line) and $k_y$ (blue line). The theoretical prediction is given by the solid black line. The dashed black line denotes the calculated shot noise standard deviation using Eq. \ref{eq:Fourier-shot-noise-variance}.}
    \label{fig:PSF-MTF-full}
\end{figure}

\begin{figure}
    \centering
    \includegraphics[scale=0.335]{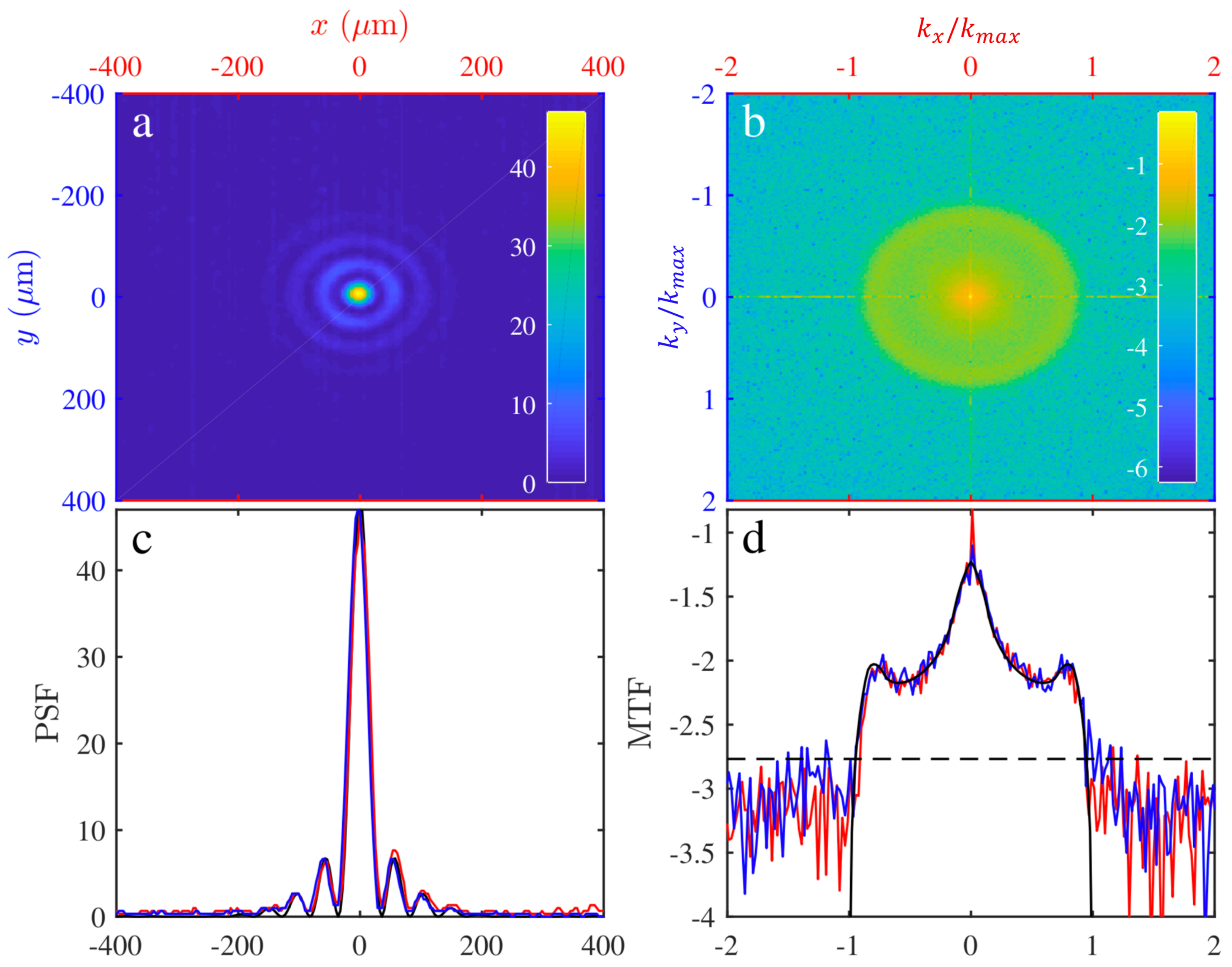}
    \caption{Measured PSF and MTF for annular pupil with $k_-=k_{max}/2\sqrt{2}$, NA$=0.0066$, and exposure time of $T=4$ s. The subfigures are defined in the same manner as Fig. \ref{fig:PSF-MTF-full}.}
    \label{fig:PSF-MTF-annular}
\end{figure}

Examples of the experimentally measured PSFs and MTFs are presented in Figs. \ref{fig:PSF-MTF-full} and \ref{fig:PSF-MTF-annular} for the open and annular pupils, respectively. Also shown are the theoretical PSFs and MTFs determined from scalar diffraction theory.
A good agreement can be seen with both the PSF and MTF between theory (black lines) and experiment (red and blue lines), indicating that the imaging system is well aligned and not inducing any unwanted aberrations. Also, the calculated standard deviation (black dashed line) seems to accurately predict where the MTF is overcome by the shot noise, providing evidence supporting the theoretical model for spectral noise. Since $\sigma_e=0$ in this calculation, it is apparent that the readout noise is in fact likely negligible.

\subsection{Spectral SNR}

\begin{figure}
    \centering
    \includegraphics[scale=0.28]{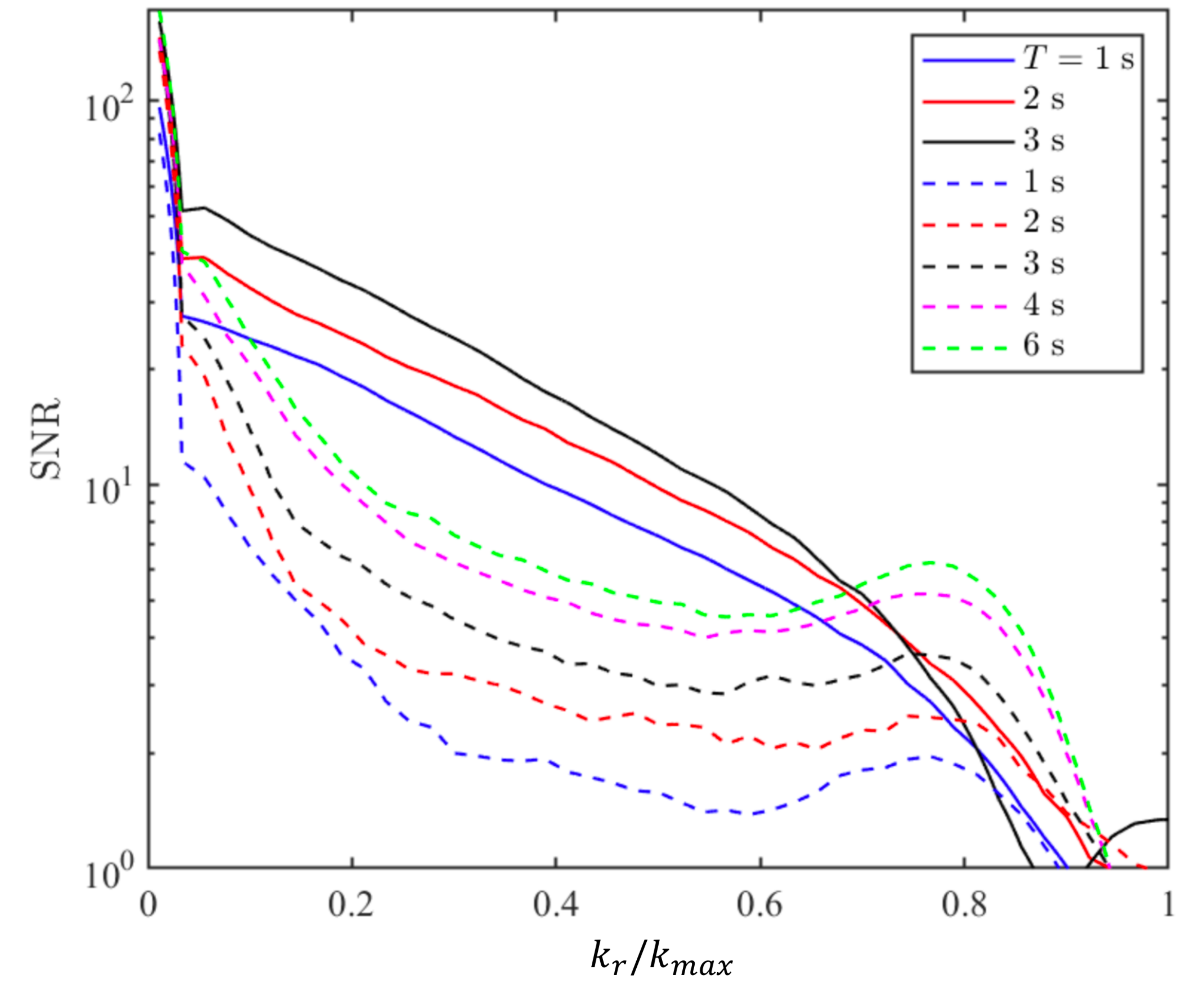}
    \caption{Measured spectral SNR for open (solid lines) and annular (dashed lines) pupils for different exposure times. The crossover by the black solid line over the red and blue lines is due to distortion of the PSF by pixel saturation at $T=3$ s for the open pupil.}
    \label{fig:spectral-SNR}
\end{figure}

From the OTFs computed from the measured PSFs in Figs. \ref{fig:PSF-MTF-full} and \ref{fig:PSF-MTF-annular}, it is then straightforward to compute $\mathrm{SNR}_q$ for each exposure using Eq. \ref{eq:spectral-SNR}, assuming that the experimental pixel values and the number of photons at each pixel are about linearly related. These are plotted in Fig. \ref{fig:spectral-SNR}, where the solid lines indicate the open pupil $\mathrm{SNR}_q$ and the dashed lines indicate the annular pupil $\mathrm{SNR}_q$.

\begin{figure}
    \centering
    \includegraphics[scale=0.27]{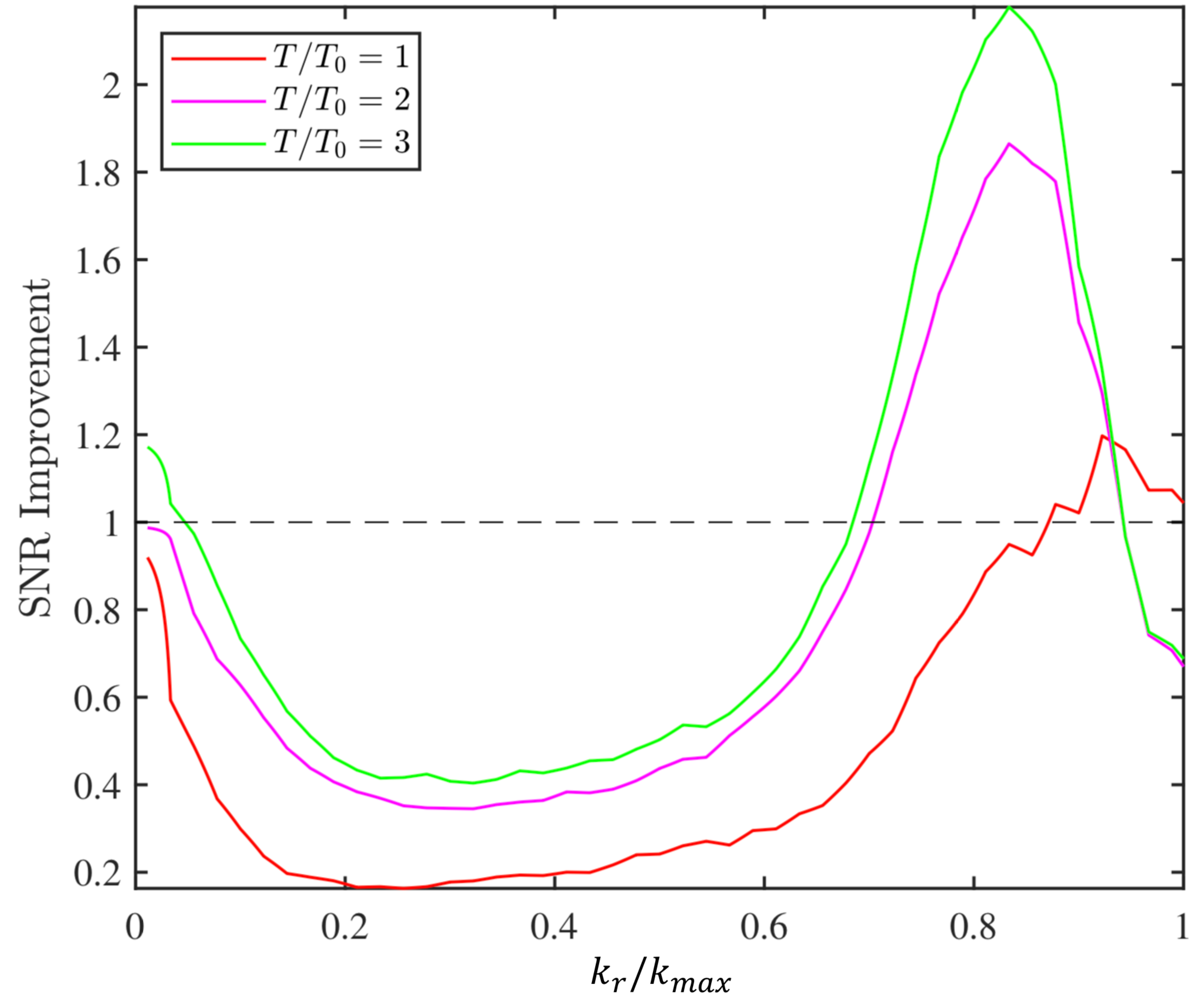}
    \caption{Spectral SNR improvement $\text{SNRi}_q$ corresponding to the data in Fig. \ref{fig:spectral-SNR} with $T_0=2$ s. The black dashed line indicates $\text{SNRi}_q=1$.}
    \label{fig:spectral-SNR-improvement}
\end{figure}

For a direct comparison of the $\mathrm{SNR}_q$ for the two pupil configurations, we define a spectral SNR improvement metric
\begin{equation}
    \text{SNRi}_q=\frac{\mathrm{SNR}_{q,a,T}}{\mathrm{SNR}_{q,o,T_0}},
\end{equation}
where $\mathrm{SNR}_{q,a,T}$ is the annular pupil spectral SNR for exposure time $T$ and $\mathrm{SNR}_{q,o,T_0}$ is the open pupil spectral SNR for exposure time $T_0$. Plotted in Fig. \ref{fig:spectral-SNR-improvement} is the $\text{SNRi}_q$ for $T_0=2$ s and three different values of $T$.
It can be seen that a clear enhancement in the SNR for spatial frequencies near $0.83k_{max}$ can be obtained using an annular pupil provided a sufficient exposure.

\subsection{Test Images}

\begin{figure}
    \centering
    {\includegraphics[width=0.98\textwidth]{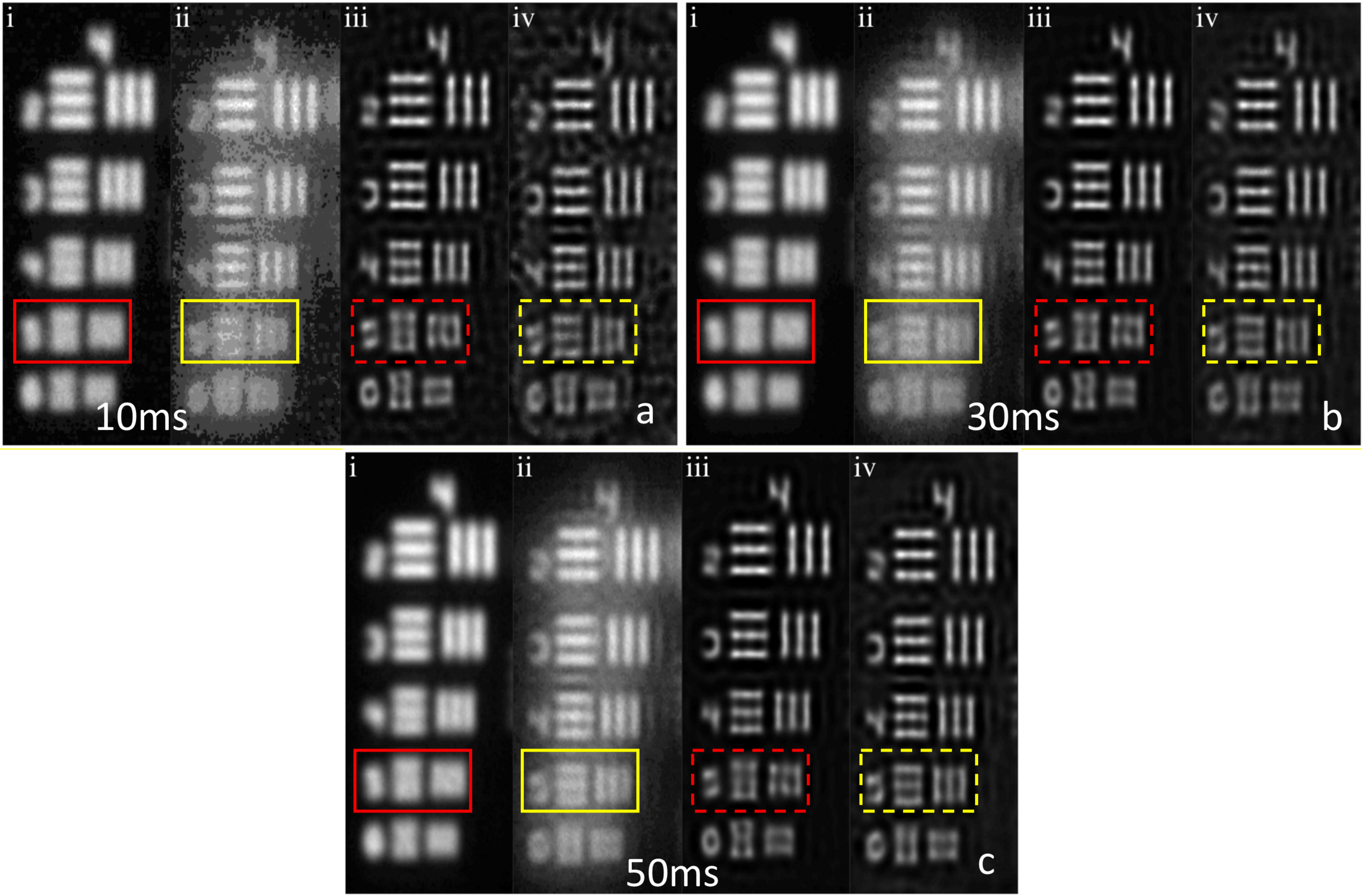}}
    \caption{Experimental images of a USAF-1951 resolution test target collected from the setup in Fig. \ref{fig:experiment-diagram} with (a) $T=10$ ms, (b) $T=30$ ms, and (c) $T=50$ ms exposure times. The individual images in each subfigure correspond to the following: i. Raw image collected with open pupil in the Fourier plane and NA=$0.0066$. ii. Raw image with annular pupil in the Fourier plane and same NA as i. iii. Image from part i deconvolved by the Richardson-Lucy algorithm after 30 iterations. iv. Image from part ii deconvolved by the Richardson-Lucy algorithm after 30 iterations. Red rectangles highlight the raw (solid) and deconvolved (dashed) images for element 5 using the open pupil. Similarly, the yellow rectangles do so for the annular pupil. Clear improvement in image resolution and contrast is seen with the annular pupil. The ACI here is implemented with annular pupil, and the reference system with open pupil.}
    \label{fig:experimental-images}
\end{figure}

To verify that the high spatial frequency improvement in spectral SNR with sufficient exposure manifests as improved image resolution, we imaged a USAF-1951 resolution test target (Thorlabs R1DS1N) using the setup in Fig. \ref{fig:experiment-diagram}. The collected images for the open and annular pupil are shown in Fig. \ref{fig:experimental-images} for three values of $T$.
Also shown is the corresponding reconstructions obtained by deconvolving the images with the Richardson-Lucy algorithm \cite{richardson1972bayesian,lucy1974iterative} as implemented in MATLAB. In Fig. \ref{fig:experimental-images}, it can be seen that three-bar target in element 5 is always blurred together by the open pupil (red solid rectangles), however even for low exposure (e.g. $T=10$ ms), the bars become qualitatively visible in the annular pupil case (yellow solid rectangles). After reconstruction for 30 iterations, element 5 remains unresolved in the open pupil images (red dashed rectangles), but the annular pupil images of element 5 are further improved (yellow dashed rectangles). 

When the exposure time $T$ is increased, the spectral SNR for the experimental MTF (see Figs. \ref{fig:PSF-MTF-full} and \ref{fig:PSF-MTF-annular}) is scaled by a factor of $\sqrt{T}$ (see Eqs. \ref{eq:photons-at-pth-pixel} and \ref{eq:spectral-SNR}) as long as no pixel saturation takes place. Particularly, as shown in Figs. \ref{fig:spectral-SNR} and \ref{fig:spectral-SNR-improvement}, the annular pupil selectively amplifies the amplitudes of the spatial frequencies within a specific range and improves the spectral SNR in that region even beyond the point which may not be otherwise possible  due to the pixel saturation with the open pupil (i.e., the reference system). This enhancement may occur at the expense of reduced spectral SNR at the other regions (e.g., lower spatial frequencies). However, those regions may be recovered easily with deconvolution. Particularly for elements 2-4 in Fig. \ref{fig:experimental-images}, the annular pupil image is improved greatly by the reconstruction since those spatial frequencies were originally attenuated with respect to the open pupil. Importantly, since the spectral noise variance is equal to the total number of photons in the entire image (see Eq. \ref{eq:Fourier-shot-noise-variance}), it is indeed advantageous to suppress the energy-rich regions so that the selective amplification successfully delivers power to the most demanding spectral band without (excessively) amplifying the noise. As a result, such a noise behavior driven spectrum manipulation, critical to the implementation of the ACI, enables us to push the SNR limit for the reference imaging system and significantly enhance the overall image contrast and resolution as can be clearly seen in Fig. \ref{fig:experimental-images}. The improvement is prominent if, for instance, element 5 (both the raw and deconvolved images) obtained from the open pupil with $T=10$ ms is compared with that of the annular pupil with $T=50$ ms. Conversely, if element 5 obtained from the open pupil with $T=50$ ms is compared with that of the annular pupil with $T=10$ ms, it is also observed that the ACI can achieve higher resolution and contrast with shorter exposure---a highly desirable feature for bioimaging modalities to avoid photo-damage to the sample.

\subsection{Pixel Saturation}

Upon viewing the open pupil $\mathrm{SNR}_q$ from Eqs. \ref{eq:photons-at-pth-pixel} and \ref{eq:spectral-SNR}, it would seem that simply increasing $T$ would lead to an improvement in $\mathrm{SNR}_q$ itself, without having to modify the pupil. However, the pixels in typical digital cameras only have finite well depth and dynamic range, meaning they can experience saturation for long exposures and/or intense illumination. The saturation causes a nonlinear response of the pixel as a function of the input photon signal. Therefore, one cannot arbitrarily increase $T$ or illumination intensity to increase $\mathrm{SNR}_q$. In terms of spatial resolution, saturation can manifest as an effective blurring due to clipping of the pixel values and blooming of photoelectrons to adjacent pixels.

\begin{figure}
    \centering
    \includegraphics[scale=0.19]{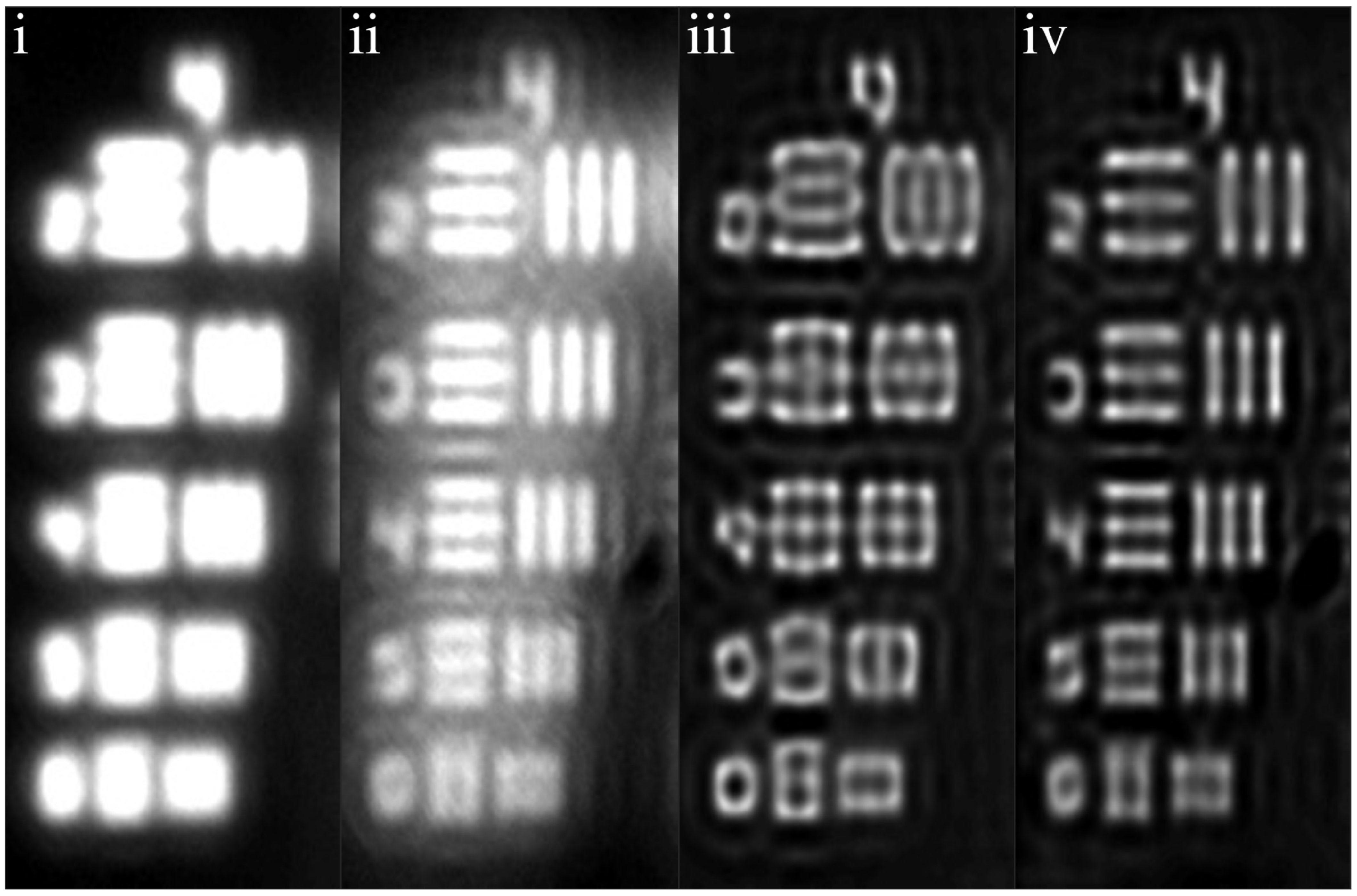}
    \caption{Images demonstrating the resistance of annular pupil to deleterious effects caused by detector saturation. Parts i-iv are defined similar to Fig. \ref{fig:experimental-images}. For the open pupil, $T=150$ ms, and for the annular pupil, $T=300$ ms. It can be seen that even for twice as long exposure time, the annular pupil image quality is mostly maintained compared to the open pupil image, which is severely blurred due to pixel saturation.}
    \label{fig:imaging-saturated}
\end{figure}

Along with provided improved resolution and contrast, the proposed spectrum manipulation method can also provide resistance to pixel saturation in cases when long exposure or intense illumination \textit{and} high resolution is required. To demonstrate this, we collected images in which the pixels become saturated for the open pupil system, and compared them to images in the annular pupil system. They are shown in Fig. \ref{fig:imaging-saturated}.
Clearly, the annular pupil has higher resistance to pixel saturation than the open pupil, despite twice longer exposure time. Element 5 can never be resolved with the open pupil as the pixel saturation appears earlier. A larger portion of the Fourier plane is blocked by the annular pupil. As evidenced by Fig. \ref{fig:pupil-functions}, the blocked photons correspond to lower spatial frequencies which are more likely to contribute to pixel saturation, since the transmission for these portions of the object will be high due to the larger local photon flux. Therefore, a cleverly tuned selective amplification does not only improve the resolution and contrast (see Fig. \ref{fig:experimental-images}), but also renders the imaging system more immune to the pixel saturation (see Fig. \ref{fig:imaging-saturated}).



\section{Discussion}

In this work, active convolved illumination, a versatile resolution enhancement method, for far-field imaging systems was presented and demonstrated experimentally for the first time. This method utilizes manipulation of the Fourier plane in the imaging system, based on an auxiliary that relies on the noise behaviour and correlates with the object pattern in order to tame the image spectrum SNR. Here, this is simply achieved by an embedded auxiliary and an annular pupil with a variable exposure time. Placing an annular pupil transparency in the Fourier plane while adjusting the photon exposure leads to improved SNR for larger spatial frequencies, which gives improved resolution over an imaging system with unobstructed pupil and the same numerical aperture. Also, the annular pupil increases resistance to pixel saturation, since a portion of the Fourier plane is blocked and the number of photons at the detector is decreased. The theory and simulation of the proposed loss mitigation and resolution enhancement method for incoherent light, the experiment to verify and expand upon the previous metamaterial implementations at near-field \cite{sadatgol2015plasmon,ghoshroy_active_2017,adams_plasmonic_2018,ghoshroy_hyperbolic_2018,ghoshroy_enhanced_2018,ghoshroy2020theory}, and how to manipulate the pupil or OTF of an imaging system to achieve a desired high spatial frequency SNR have been discussed.

It is worth mentioning that the implementation of ACI presented here should not be confused with incoherent edge detection as the latter usually relies on bipolar incoherent image processing \cite{bhuvaneswari2014edge,lohmann1978two,zhang2020sectioning}. Also, although some of our methodologies are inspired by the recent split-pupil-optimization technique \cite{becker_better_2018}, the proposed technique here is more general, does not contend with a constant ``photon budget,'' and is intimately connected with an earlier theoretical proposal \cite{sadatgol2015plasmon} as illustrated schematically in Fig. \ref{fig:ACI} (b). The ACI offers new opportunities for not only conventional imaging systems and superlenses, but also various other linear systems. One can envision its potential generalization and ubiquity to encompass imaging through random media (e.g.., turbulent and scattering atmosphere) \cite{andrews2005laser,eyyubouglu2008scintillation,hanafy2014detailed,hanafy2015estimating,hanafy2015reconstruction}, spectroscopy \cite{guerboukha1018toward,ahi2019a}, ${\cal PT}$-symmetry \cite{monticone2016parity,ganainy2019dawn,li2020virtual}, photolithography \cite{gao2015enhancing,liang2018achieving}, and quantum information and image processing \cite{defienne2019quantum,mikhalychev2019efficiently,gregory2020imaging,yuan2016quantum}. The virtual gain mechanism employed in ACI can alleviate the stringent requirement of balance between loss and gain in ${\cal PT}$-symmetric systems \cite{li2020virtual}. It is important to note that the systems with more intricate noise/distortion characteristics (e.g., turbulent atmosphere) may necessitate more ingenious engineering of the pupil function than here. Optimal and precise manipulation of the pupil function, beyond conventional optics, may be possible with the advent of amenable metasurfaces \cite{yu2014flat,zhou2019multifunctional,zhou2020flat,rho2020metasurfaces}. More work on ACI and potential relevance with structured illumination microscopy \cite{schermelleh2019super,kner2009super,schermelleh2008subdiffraction,york2012resolution,york2013instant,ingerman_signal_2019}, superoscillatory imaging \cite{yuan2019far,pu2019unlabelled,pu2020label,chen2019superoscillation,yuan2016quantum}, and super-gain \cite{di1952super,di1956directivity} can enrich further understanding of imaging beyond the known boundaries. Computational superresolution techniques \cite{wang2019deep,cox1986information,katznelson2004introduction,bertero2003super,zeng2017computational,xing2020computational} integrated with ACI can provide extended resolution limits. Machine learning approaches \cite{wang2019deep} bolstered with ACI may enable optical visualization of dynamic scenes at previously inaccessible scales. An interesting direction of research could be utilizing the analytical continuity principle \cite{cox1986information,katznelson2004introduction,wang2019deep} while enhancing the SNR with ACI to achieve far-field super-resolution imaging. Since ACI is fundamentally a physical process and enhances data acquisition, it may benefit numerous imaging scenarios. It may also be possible to use cheap detectors and yet obtain the same imaging quality as that of expensive ones not aided by ACI.

\bigskip
\noindent
\textbf{Funding.} Office of Naval Research (award N00014-15- 1-2684).



\noindent\textbf{Competing interests.} The authors declare no competing interests.

\bibliographystyle{unsrt}
\bibliography{Chapter5_v3.bib}

\end{document}